\documentclass[twocolumn,showpacs,preprintnumbers,amsmath,amssymb,footinbib]{revtex4-1}
\usepackage{graphicx}% Include figure files
\usepackage{dcolumn}% Align table columns on decimal point
\usepackage{bm}% bold math
\usepackage{mathrsfs}
\usepackage{multirow}
\usepackage[table]{xcolor}  
\begin{document}

\title{Photon pair generation by intermodal spontaneous four wave mixing in birefringent, weakly guiding optical fibers}

\author{K. Garay-Palmett$^2$, D. Cruz-Delgado$^1$, F. Dominguez-Serna$^2$, E. Ortiz-Ricardo$^1$, J. Monroy-Ruz$^1$, H. Cruz Ramirez$^1$,  R. Ramirez-Alarcon$^1$, and A. B. U'Ren$^1$}
\affiliation{$^1$Instituto de Ciencias Nucleares, Universidad Nacional Aut\'onoma de M\'exico, apdo. postal 70-543, 04510 D.F., M\'exico\\ $^2$Departamento de \'Optica, Centro de Investigaci\'on Cient\'ifica y de Educaci\'on Superior de Ensenada, Apartado Postal 360 Ensenada, BC 22860, M\'exico}
\date{\today}

\begin{abstract}
We present a theoretical and experimental study of the generation of photon pairs through the process of spontaneous four wave mixing (SFWM) in a few-mode, birefringent fiber.   Under these conditions, multiple SFWM processes are in fact possible, each associated with a different combination of transverse modes for the four waves involved.  We show that in the weakly guiding regime, for which the propagation modes may be well approximated by linearly polarized modes, the departure from circular symmetry due to the fiber birefringence translates into conservation rules which retain elements from azimuthal and rectangular symmetries: both OAM and parity must be conserved for a process to be viable.     We have implemented a SFWM source based on a ``bow-tie'' birefringent fiber, and have measured for a collection of pump wavelengths the SFWM spectra of each of the signal and idler photons in coincidence with its partner photon.  We have  used this information, together with knowledge of the transverse modes into which the signal and idler photons are emitted, as input for a genetic algorithm which accomplishes two tasks: i) the identification of the particular SFWM processes which are present in the source, and ii) the characterization of the fiber used. 
\end{abstract}

\maketitle

\section{Introduction}

The process of spontaneous four wave mixing (SFWM)~\cite{fiorentino02} has, over the last decade and a half, become a viable alternative based on the $\chi^{(3)}$ non-linearity of optical fibers for the generation of photon pairs, to the more established process of spontaneous parametric downconversion (SPDC) in $\chi^{(2)}$ non-linear crystals~\cite{Burnham70}.    The implementation of SFWM sources with fibers which support more than one transverse mode leads to a wealth of possibilities, some of which are explored in this paper.    

The transverse modal content of a fiber can be exploited for increasing the transmission capacity in optical communication systems \cite{randel11},  and in nonlinear optics may be used for tuning the frequencies of operation \cite{chen13,shavrin13}. In the context of $\chi^{(2)}$ non-linear waveguide SPDC sources, the use of transverse modes has been explored in a number of papers~\cite{mosley09,christ09,kruse13,karpinski09,saleh09}.  Transverse spatial modes represent an essential feature of optical fibers; in this paper, we focus on the interplay of the spectral and transverse spatial mode degrees of freedom in the SFWM process.    We show that the two-photon state obtained through SFWM in a fiber which supports more than one transverse mode in general exhibits hybrid entanglement in frequency and transverse mode~\cite{neves09}.  Importantly, spatial entanglement based on transverse fiber modes is scalable to higher dimensions as controlled by the number of supported fiber modes, unlike polarization entanglement which is limited to a dimension of $2$.   A highly important feature of higher-order fiber modes is that they may carry orbital angular momentum (OAM), so that  transverse mode entanglement can imply the presence of OAM entanglement. 

Building on previous work from our group~\cite{cruz14},  we have concentrated on the use of few-mode, weakly guiding, bi-refringent fibers~\cite{smith09,zhou13,fang14,fan07} as a useful experimental platform for the study of two-photon states which can exhibit hybrid entanglement in frequency and transverse mode. In such fibers, while all six electromagnetic field components of the supported modes are non-zero, they may be well approximated by linearly polarized (LP) modes. If multiple transverse modes are supported, a number of SFWM processes are possible each involving a distinct combination of transverse modes for the four participating waves: pump 1, pump 2, signal, and idler.  We discuss how in the absence of full circular symmetry the supported modes may be well-described by linearly polarized (LP) modes with well-defined even/odd parities.      It is known that  while OAM conservation is expected for circularly-symmetric guided-wave SPDC and SFWM sources, this conservation rule becomes parity conservation for sources with rectangular symmetry.  We show that SFWM in birefringent fibers retain elements from both azimuthal and Cartesian symmetries: i.e. both OAM and parity are conserved.   For a fiber which supports $M$ modes, there may be up to $M^4$ intermodal SFWM processes; we describe how parity and OAM conservation rules define which of these $M^4$ processes can actually take place.

We present experimental data of SFWM spectra obtained for a birefringent ``bow tie'' fiber, for a number of different pump wavelengths. This data shows evidence of three separate SFWM processes, manifested by three pairs of energy-conserving peaks. An analysis of the processes which conserve both OAM and parity, together with a genetic algorithm which takes the measured spectra along with the known propagation modes corresponding to each of the six peaks as input, enables us to identify the specific SFWM processes present in our source, and also yields a characterization of the fiber used.

\section{Intermodal spontaneous four wave mixing in birefringent fibers} 

\subsection{The two-photon state for multiple intermodal SFWM processes}

In an optical fiber which supports more than one transverse mode, several intermodal SFWM processes can take place. The resulting  two-photon state is then a coherent superposition of the contributions from the different processes, each of which is associated with a particular combination of transverse modes for the four waves involved (pump 1, pump 2, signal, and idler)~\cite{cruz14}.  Here we will consider experimental situations where while the two pumps may be non-degenerate in transverse mode, they are spectrally degenerate.

As will be discussed in detail below, among all possible combinations of supported modes for the four waves, those that i) are phasematched, and ii) have a non-zero transverse mode overlap, lead to allowed SFWM processes.  If there are $N$ such allowed processes (in the spectral range of interest), the two photon state in general exhibits hybrid entanglement in frequency and in transverse mode, and is 
given by $|\Psi\rangle=|\mbox{vac}\rangle+\eta |\Psi_2\rangle$, with

\begin{align}
\label{eq:state}
|\Psi_2\rangle &= \sum_{j=1}^N \sqrt{W_{j1} W_{j2}}\mathscr{O}_j(\alpha_j,\beta_j,\mu_j,\nu_j) \\ \nonumber
&\times \int\!\!d\omega_s\!\!\int\!\!d\omega_i\, f_j(\omega_s,\omega_i)\hat{a}^\dag(\omega_s;\mu_j)\hat{a}^\dag(\omega_i;\nu_j)|\mbox{vac}\rangle,
\end{align}

\noindent where for process $j$, $\hat{a}^\dag(\omega_s;\mu_j)$ ($\hat{a}^\dag(\omega_i;\nu_j)$) is the creation operator for the  signal (idler) mode with frequency $\omega_s$ ($\omega_i$) propagating in transverse spatial mode $\mu_j$ ($\nu_j$), and where the pump 1 (pump 2) wave propagates in transverse mode $\alpha_j$ ($\beta_j$) with power $W_{j1}$ ($W_{j2}$).  
$\mathscr{O}_j(\alpha_j,\beta_j,\mu_j,\nu_j)$ is the transverse mode overlap between the four waves, expressed in terms of the transverse electric field distribution $g(\xi; \bm{\rho}^\bot )$, for transverse mode $\xi$ and dependent on the the transverse position $\bm{\rho}^\bot$, as

\begin{align}
\label{eq:overlap}
\mathscr{O}_j(\alpha_j,\beta_j,\mu_j,\nu_j)=M_j  \int &d^2 \bm{\rho}^\bot  \ g(\alpha_j; \bm{\rho}^\bot ) g(\beta_j; \bm{\rho}^\bot ) \\ \nonumber&\times g^*(\mu_j; \bm{\rho}^\bot ) g^*(\nu_j; \bm{\rho}^\bot ).
\end{align}

Note that in writing equation (\ref{eq:overlap}), we have assumed that the dependence on frequency for each of the transverse electric field distributions $g(\xi; \bm{\rho}^\bot )$ may be neglected. Also, $M_j$ is a normalization constant chosen so that the sum over all $j$ of $|\mathscr{O}_j(\alpha_j,\beta_j,\mu_j,\nu_j)|^2$ yields unity.

For process $j$, $f_j(\omega_s,\omega_i)$ represents the joint spectral amplitude (JSA) which is determined by the phasematching characteristics and is given by~\cite{garay07}

\begin{align}
\label{eq:jsa}
f_j(\omega_s,\omega_i)=\int\!\!d\omega\,A(\omega)A(\omega_s+\omega_i-\omega)\mbox{sinc}\left[\frac{L}{2}\Delta k_j  \right],
\end{align}

\noindent written in terms of the fiber length $L$ and  the (degenerate) pump spectral envelope $A(\omega)$;  the function 
 $I_j(\omega_s,\omega_i)=|f_j(\omega_s,\omega_i)|^2$ is referred to as the joint spectral intensity (JSI).
In equation (\ref{eq:jsa}), $\Delta k_j$ represents the phase mismatch given by

\begin{align}
\label{eq:DK}
\Delta k_j&=k(\omega;\alpha_j)+k(\omega_s+\omega_i-\omega;\beta_j)\\ \nonumber&-k(\omega_s;\mu_j)-k(\omega_i;\nu_j)-\phi_{NLj}
\end{align}

\noindent  given in terms of the wavenumber $k(\omega;\xi)$ for transverse mode $\xi$ and frequency $\omega$; $\phi_{NLj}$ is a non-linear contribution determined by self-and cross-phase modulation~\cite{garay07}.

Note that while a number of works have studied in detail the longitudinal phasematching properties, as defined by the JSA function (see Eq.~\ref{eq:jsa}), in the theoretical part of this paper we focus on the transverse mode overlap term $\mathscr{O}_j(\alpha_j,\beta_j,\mu_j,\nu_j)$ and on the new physics which may be derived henceforth.

While the theory presented so far is general and can be applied to any type of fiber, we are interested in particular in SFWM sources implemented with birefringent fibers.    
Knowledge of the modes which are supported by the SFWM fiber is required for a theoretical description of the two-photon state.  Our work, presented below, is based on a fiber in which the circular symmetry of the fiber is (slightly) broken by the birefringence introduced by two stress rods on either side of the core.   In subsection B, below, we discuss how the fiber modes are well described by linearly-polarized (LP) modes, appropriately modified by polarization and parity.   

Let us note that if a single SFWM process (corresponding to a particular value of $j$ in Eq. \ref{eq:state}) were to be postselected, then the resulting two-photon state can be entangled in frequency but is otherwise unentangled; in particular, since the post-selected signal and idler photons  propagate in known transverse modes  $\mu_j$ and $\nu_j$, spatial and mixed spatial-spectral entanglement is suppressed.    Each process has an efficiency defined by the overlap between the four waves involved $\mathscr{O}_j(\alpha_j,\beta_j,\mu_j,\nu_j)$.    For $M$ supported modes, we can have in principle up to $M^4$ SFWM processes, so that an important question which we address in this paper is which of these $M^4$ processes actually take place in a given situation.   We will show for the particular case of linearly polarized (LP) modes, that this mode overlap is non-zero if both OAM and parity are conserved.    This is crucial for an understanding of the two-photon source,  because it enables the identification of the processes which are viable.   It is important to point out that the number of viable processes can be drastically reduced by OAM and parity conservation from the $M^4$ possible processes.

We point out that phasematching properties are such that typically signal and idler emission frequencies are correlated to the particular spatial modes involved in each process.   This implies that the postselection referred to in the previous paragraph can be accomplished in a straightforward manner through spectral filtering.   If, however, no postselection is carried out then the two-photon state is given by a coherent sum as indicated in Eq. \ref{eq:state}.    In this case, the two-photon state can exhibit hybrid frequency-transverse mode entanglement.    While OAM and parity for the overall two-photon state are not well-defined, and therefore it is unclear how to formulate OAM and parity conservation rules for this overall state, these quantities must be conserved, as we show in subsection C below, on a \emph{process-by-process} basis.

In this paper, we concentrate on cross-polarized SFWM sources implemented with few-mode, weakly guiding, bi-refringent fibers.   The essential advantage of such SFWM sources is that the fiber birefringence results in a shift  (with respect to a co-polarized SFWM process in non-birefirngent fiber) in the phasematching condition leading to signal and idler frequencies $\omega_s$ and  $\omega_i$ which are sufficiently removed from the pump frequency $\omega_p$  to avoid Raman contamination, while group velocity matching conditions can be fulfilled permitting various types of engineered spectral correlations \cite{cruz15}.    Such fibers have indeed been used as the basis for a number of recent experiments~\cite{smith09,zhou13,fang14,fan07} .  In this paper we aim to provide a complete theoretical framework which permits the full description of the SFWM two-photon state produced in such fibers; such a complete theoretical framework has not been presented in the literature.   This framework must include reliable knowledge of the transverse modes involved, incorportating the effect of parity and birefringence, as will be discussed in subsection B.  An important aspect, discussed in subsection C, is how  parity and OAM conservation can help define which SFWM processes actually take place.   The dispersion model which we have used, including polarization and parity birefringence, required for carrying our specific simulations of the two-photon state, is described in section D.  

\subsection{Transverse modes in weakly-guiding, birefringent fibers} \label{modeTh}

A conventional optical fiber exhibits a step refractive index profile. It comprises a core with radius $r_0$ and refractive index $n_1$, and a cladding with refractive index $n_2$ ($n_1>n_2$); 
the electromagnetic formalism for such fibers is well known~\cite{snyder78}.   While the exact propagating modes are hybrid in the sense that all six components of the electromagnetic field are non-zero, for fibers characterized by a low dielectric contrast  ($n_1-n_2\ll1$), it is well known that the propagating modes are approximately linearly polarized. Thus, for circularly-symmetric fibers with a low dielectric contrast, transverse modes are well-described by the  $\mbox{LP}_{lm}$ family of modes, where $l=0,1,2,\ldots$ and $m=1,2,\ldots$ indicate the number of maxima of the spatial profile along the azimuthal and radial directions, respectively. Note that the mode index $l$ is related to the orbital angular momentum (OAM) of light; in particular, a  mode with subindex $l$ corresponds to a linear combination of contributions with azimuthal dependence $\exp(\pm i l \phi)$, i.e. with  topological charge $\pm l$.  It can be shown that an $\mbox{LP}_{lm}$ mode arises from the superposition of the nearly-degenerate exact modes $\mbox{HE}_{l+1,m}$ and $\mbox{EH}_{l-1,m}$; while neither of these modes is linearly polarized, the sum is in fact nearly linearly polarized~\cite{snyder78}.   

In our work we are interested in birefringent fibers which depart from being circularly symmetric.   Let us note that for circularly symmetric fibers, we are free to express the azimuthal dependence of the LP modes in an exponential $\{\exp( i l \phi), \exp(- i l \phi)\}$ or in a  $\{\mbox{sin}(l \phi), \mbox{cos}(l \phi)\}$ basis.   However, in the presence of fiber birefringence for which azimuthal symmetry is broken,  modes have a well-defined even/odd parity and modes with an exponential  $\exp(\pm i  l \phi)$ azimuthal dependence  are in fact not supported.  Thus, for a birefringent fiber the propagation modes are correctly expressed in the sine/cosine basis, with an appropriate parity-dependent propagation constant.   This is consistent with the fact that the SFWM photons emitted by our source, see the experimental section below, indeed have well-defined parities.

\begin{figure}
\includegraphics[width=8.4cm]{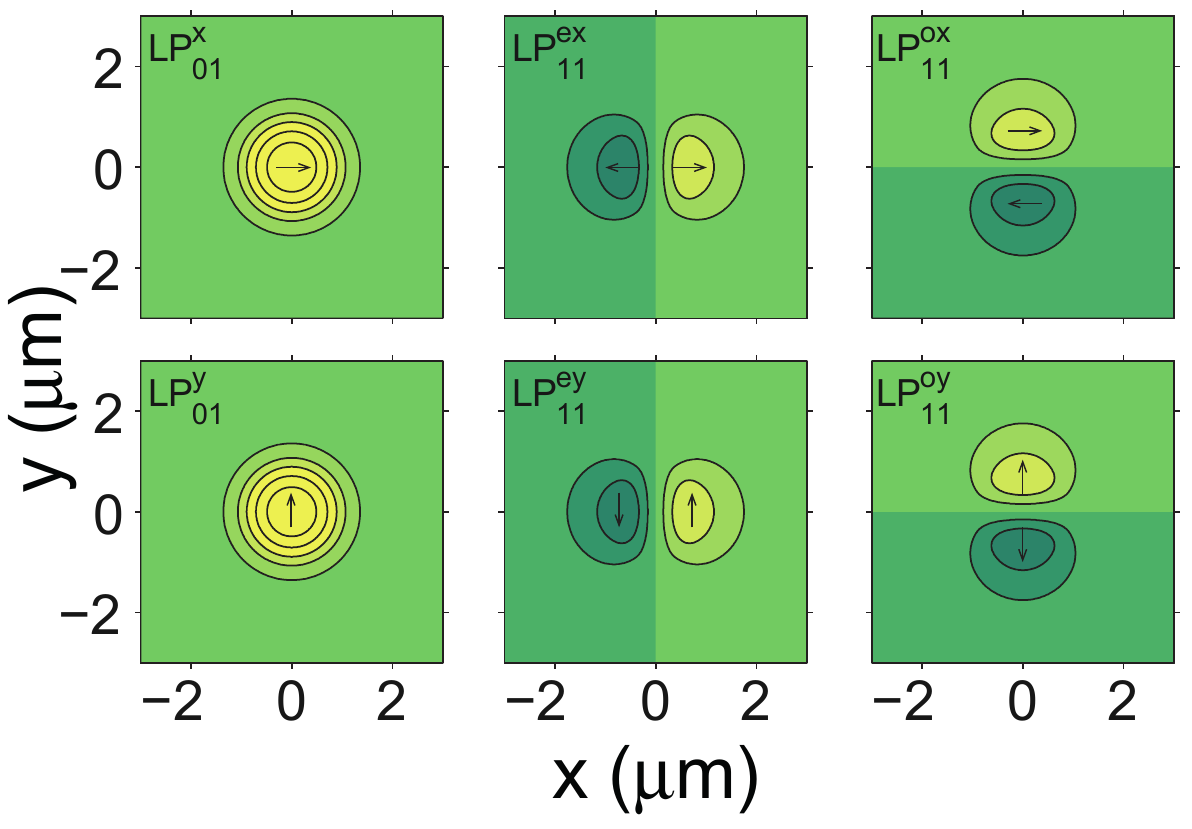}
\caption{\label{fig:LP} Unfolding of the $LP_{01}$ and  $LP_{11}$ modes.}
\end{figure}

Each propagation mode has two additional properties, besides the number of radial and azimuthal maxima, which are essential in our analysis: polarization and parity.   In fact, the fundamental mode   LP$_{01}$ actually represents two modes,  the $x$-polarized LP$^x_{01}$
and the $y$-polarized LP$^y_{01}$  modes, which become degenerate (i.e. have the same propagation constant) in the case of circularly-symmetric fibers.   For higher-order modes, parity becomes important~\cite{snyder78,kim87}, so that for example the mode LP$_{11}$ actually represents the four modes  LP$^{ex}_{11}$, LP$^{ey}_{11}$, LP$^{ox}_{11}$, and LP$^{oy}_{11}$, where $e$ ($o$) signifies even (odd) parity.   Again, while these four modes are degenerate for circularly-symmetric fibers, in the case of a birefringent fiber,  the two  LP$_{01}$ and the four LP$_{11}$ modes become non-degenerate and must be taken into account separately.  A graphical summary of this ``unfolding'' of the  LP$_{01}$ and LP$_{11}$ modes into six separate modes is shown in  figure \ref{fig:LP}.  Note that all modes LP$_{lm}$ with $l \ge 1$ likewise unfold into four separate modes, which become non-degenerate in fibers which depart from circular symmetry.

The transverse mode set for birefringent fibers is consequently well represented  by the  linearly polarized (LP) modes, with four labels, written as  $\mbox{LP}_{lm}^{pq}$.   While  $l=0,1,2,\ldots$ and $m=1,2,\ldots$ indicate the number of maxima of the spatial profile along the azimuthal and radial directions, respectively, $p=x(y)$ indicates that the mode is polarized along the $x$ ($y$)  direction and $q=e (o)$ indicates the mode has even (odd) parity. Note that the x (y) polarization in our case is defined as parallel to the slow axis (fast axis) of the birefringent fiber used.  Note, also,  that while the parity can be even or odd for all values $l \ge 1$,  $l=0$ modes always have even parity.   The subscripts $\alpha_j,\beta_j,\mu_j,\nu_j$ in equations (\ref{eq:overlap}) and (\ref{eq:DK}) then represent particular combinations of values for the four mode indices $l$ and $m$, $p$, and $q$. 

The transverse electric field distribution for mode $\mbox{LP}_{lm}^{pq}$, expressed in polar coordinates $r$ and $\phi$,  can be factored into radial and azimuthal factors as follows

\begin{equation}
\label{eq:TFD}
g_{lm}^{pq}(r,\phi)=F_{lm}^{pq}(r) G_l^q(\phi).
\end{equation}

In equation (\ref{eq:TFD}), the radial factor is given in terms of the $l$th order Bessel function of the first kind $J_l(.)$ and the $l$th order modified Bessel function of the second kind $K_l(.)$, as

\begin{equation}
F_{lm}^{pq}(r)= \begin{cases} J_l(u_{lm}^{pq}r), & r<r_0    \\    K_l(v_{lm}^{pq}r), &   r>r_0 \end{cases}
\end{equation}

\noindent where $r_0$ is the core radius and where

\begin{align}
u_{lm}^{pq}&= \sqrt{ k_1^2-(k_{lm}^{pq})^2 } \nonumber \\
v_{lm}^{pq}&= \sqrt{ (k_{lm}^{pq})^2-k_2^2}
\end{align}

\noindent are parameters defined in terms of the wavenumbers for frequency $\omega$ of the propagation mode, $k_{lm}^{pq}$, 
as well as wavenumbers corresponding to the core, $k_{1}$, and the cladding, $k_{2}$, expressed as

\begin{align}
k_{lm}^{pq} &= n_{lm}^{pq} \omega / c  \nonumber \\
k_{1} &= n_{1} \omega / c \nonumber \\
k_{2} &= n_{2} \omega / c.
\end{align}

In the equation above, $n_{lm}^{pq} $ represents the effective index in the fiber of the propagation mode.  Note that  for the specific case $l=0$, there is no azimuthal dependence and $G_{0}^q(\phi)=1$ if $q=1$ (even parity), while $G_{0}^q(\phi)$ is undefined if $q=-1$ (odd  parity),  i.e. the modes with $l=0$ necessarily have an even parity. For $l \ge 1$,

\begin{equation}\label{eq:azimuthalfield}
G_{l}^q(\phi)=\begin{cases}\cos(l \phi), \ \ \ q=1 & (\mbox{even mode})  \\   \sin(l \phi), \ \ \ q=-1 & (\mbox{odd mode})\end{cases}.
\end{equation}

We will now proceed to investigate the form which the mode overlap (see Eq.~(\ref{eq:overlap})) takes for LP modes.

\subsection{Conservation of orbital angular momentum and parity in SFWM}\label{cons}

The question of whether OAM and parity are conserved for the spontaneous parametric downconversion (SPDC) process has been addressed by a number of authors.   OAM conservation was assumed to hold in the early work on OAM entanglement with non-guided  SPDC~\cite{mair01};  it was identified then that OAM conservation leads naturally to OAM entanglement.   It was later determined that in fact OAM conservation is only to be expected in those particular cases for which the interaction Hamiltonian is azimuthally symmetric~\cite{arnaut00,barbosa07,feng08,barbosa09}.  Thus, in a  number of non-guided source configurations including bulk-crystal type-II, as well as type-I source designs for which Poynting vector walkoff cannot be ignored (e.g. derived from strong pump focusing or a long crystal), OAM is not conserved.      In the case of guided SPDC, i.e. occurring in a nonlinear $\chi^{(2)}$ waveguide, source symmetries dictate the conservation rules~\cite{walborn05}.   While azimuthally-symmetric circular waveguides lead to an OAM conservation rule, rectangular waveguides lead to a parity, rather than OAM, conservation rule~\cite{bharadwaj15,mosley09}.    Our own SFWM source is characterized by a cylindrical fiber core with a (slight) deviation from azimuthal symmetry which translates into non azimuthally-symmetric $l \ge 1$ modes. 
It is very interesting that for this source, elements are retained from both circular and Cartersian symmetries: as we show in this section, both OAM and parity are conserved on a \emph{process-by-process} basis.  
%
%As already mentioned, conservation of OAM and parity rules are extremely helpful in order to determine which of the $M^4$ possible processes for $M$ supported fiber transverse modes are in fact viable.  

Let us note that classical, i.e. stimulated, three and four wave mixing processes are a close relative of the spontaneous four wave mixing process studied in this paper so that our discussion, including the physics of parity and OAM conservation, of course is related to the similar effects observed classically~\cite{lin81}.   In particular, recent work has addressed these issues for stimulated four wave mixing~\cite{poletti08,ding14}.  The quantum-mechanical treatment presented here is aimed at enabling the design of a new generation of fiber-based photon pair sources with hybrid entanglement in frequency and in transverse mode.

In a fiber which supports multiple transverse modes, each possible combination of transverse modes amongst the four participating waves can result in a SFWM process provided that two conditions are fulfilled:  i) the four waves are appropriately phasematched at given pump and generation frequencies  i.e. $\Delta k_j  = 0$ (see equation (\ref{eq:DK})), or at least approximately phasematched as $|L \Delta k_j| \le 2\pi$, and ii) these four waves have a non-vanishing overlap.   In a situation where there are $M$ transverse modes available for each wave, there are $M^4$ possible processes.  For example, if the $\mbox{LP}_{01}$ and $\mbox{LP}_{11}$ are supported (which become six non-degenerate modes in the case of a birefringent fiber), there are then $6^4=1296$ possible processes. We show below that the second condition, i.e. a non-vanishing mode overlap, translates for the specific case of linearly polarized modes into conservation rules which may be used to determine the processes which are greatly reduce the number of viable processes.

Let us consider the case where each wave is described by LP modes, so that for process $j$ the pump-1 (pump-2) wave  $\alpha_j$ ($\beta_j$)  becomes characterized by four indices: $l_{j1}$, $m_{j1}$, $p_{j1}$, and $q_{j1}$ ($l_{j2}$, $m_{j2}$, $p_{j2}$, and $q_{j2}$).  Likewise, the signal (idler) wave $\mu_j$ ($\nu_j$) becomes characterized by four indices $l_{js}$, $m_{js}$, $p_{js}$, and $q_{js}$ ($l_{ji}$, $m_{ji}$, $p_{ji}$, and $q_{ji}$).   The fact that for LP modes the transverse electric field distribution $g_{lm}^{pq}(r,\phi)$ is factorable into radial and azimuthal contributions, implies that the mode overlap $\mathscr{O}_j(\alpha_j,\beta_j,\mu_j,\nu_j)$ may be likewise factored into radial and azimuthal contributions, as 

\begin{equation}
\mathscr{O}_j(\alpha_j,\beta_j,\mu_j,\nu_j)=\mathscr{O}^r_j \mathscr{O}^\phi_j,
\end{equation}

\noindent in terms of  radial and azimuthal overlap integrals, $\mathscr{O}^r_j$ and $\mathscr{O}^\phi_j$,  as follows

\begin{align}
\label{eq:Ophi}
\mathscr{O}^r_j  &=\!\int\limits_0^\infty \!\! r dr  F_{l_{j1}m_{j1}}^{p_{j1}q_{j1}} (r) F_{l_{j2}m_{j2}}^{p_{j2}q_{j2}} (r) [F_{l_{js}m_{js}}^{p_{js}q_{js}}(r)]^*[F_{l_{ji}m_{ji}}^{p_{ji}q_{ji}} (r)]^* \nonumber \\
\mathscr{O}^\phi_j  &=\int\limits_0^{2 \pi} d \phi \, G_{l_{j1}}^{q_{j1}} (\phi) G_{l_{j2}}^{q_{j2}} (\phi) [G_{l_{js}}^{q_{js}}(\phi)]^*[G_{l_{ji}}^{q_{ji}} (\phi)]^*. 
\end{align}

For the analysis below, we will focus on the azimuthal overlap $\mathscr{O}^\phi_j$, which may be expressed as 

\begin{align}
\label{eq:Ophi}
\mathscr{O}^\phi_j  &=\delta_{\Delta q_j} (\delta_{\Delta l_j^{+++}}+\delta_{\Delta l_j^{-++}}+ \ldots +\delta_{\Delta l_j^{---}})
\end{align}

\noindent where the quantity within the brackets contains terms defined by all  $8$ possible combinations of signs for $l_{j2}$, $l_{js}$, and $l_{ji}$.  In Eq.~\ref{eq:Ophi}, $\delta_n$, with integer $n$, represents a one-argument Kronecker delta (which vanishes unless $n=0$, in which case the delta yields unity).    Note that in writing Eq.~\ref{eq:Ophi} we have defined a parity non-conservation parameter for process $j$, $\Delta q_j$, as follows

\begin{equation}
\Delta q_j= q_{j1}q_{j2}-q_{js}q_{ji},
\end{equation}

\noindent as well as a family of OAM non-conservation parameters for process $j$

\begin{align}
\label{eq:Deltl}
&\Delta l_j^{+++}=l_{j1}+ l_{j2}+ l_{js}+ l_{ji} \nonumber \\ 
&\Delta l_j^{-++}=l_{j1}- l_{j2}+ l_{js}+ l_{ji} \nonumber \\ 
&\ldots  \nonumber \\ 
&\Delta l_j^{---}=l_{j1}- l_{j2}- l_{js}- l_{ji},
\end{align}

\noindent  for all $8$ combinations of signs for $l_{j2}$, $l_{js}$, and  $l_{ji}$.

It is clear from Eq.~\ref{eq:Ophi} that in order for the azimuthal overlap to be non-zero for a given process $j$, the following two conditions must be observed: i) $\Delta q_j=0$ and ii)   at least one of the eight terms within the brackets, each corresponding to a different combination of signs in front of the topological charges for three of the waves, must be non-zero, leading to the condition that at least one $\Delta l_j$ parameter must vanish.

The first of the above conditions tells us that the mode overlap vanishes unless $q_{j1} q_{j2}=q_{js} q_{ji}$, i.e. the parity of the pump waves must match the parity of the generated SFWM photons, which is a statement of parity conservation.  

In order to interpret physically the second condition, let us consider the $\mbox{LP}_{lm}^{pq}$  modes, expressed in the sine/cosine basis which is consistent with the loss of circular symmetry in our birefringent fiber, see Eqns. \ref{eq:TFD} and \ref{eq:azimuthalfield}.  Note that while these modes do not contain a phase singularity and therefore do not carry OAM, they are in fact, for $l \ge 1$,  the coherent addition of two subjacent optical vortices with topological charges $l$ and $-l$.    The second of the above conditions is then fulfilled if the sum of the topological charge values for all four waves, with any combination of signs in front of each one, is zero.   In other words, this second condition is fulfilled if OAM is conserved, i.e. the sum of the topological charges for the two pumps matches the sum of the topological charges for the two generated photons, 
when any of the two subjacent vortices is selected for each of these four waves.

It is important to point out that the parity and OAM conservation rules presented here cannot apply to the overall two-photon state; indeed, it is not clear how to define parity and OAM for the overall two-photon state.   Thus, parity and OAM conservation rules apply on a \emph{process by process} basis, and determine whether a particular SFWM process $j$ is viable.

\subsection{Simplified model for dispersion properties of birefringent fibers}

Birefringent fibers allow cross-polarized  phasematching in SFWM \cite{smith09,meyer13,zhou13,fang14,cruz14}. In this paper, we focus on processes of the kind $xx-yy$, in which the pump and the emitted photons have orthogonal polarizations.  While fiber birefringence is essential in our experimental implementation so as to ensure phasematching, it is in fact important in other ways.   Indeed, such a cross-polarized scheme simplifies the separation of SFWM from pump photons at the fiber output, and contributes to minimize the noise background produced by spontaneous Raman scattering \cite{lin07}.    For an optimal implementation of this configuration it is desirable that the state of polarization of the waves involved  remain unchanged along the fiber, which becomes viable in highly birefringent fibers, known as polarization maintaining fibers, for which the difference in propagation constant for the two orthogonal polarizations is obtained by designing elliptical cores or including ``panda''- or ``bow-tie''-type stress rods into the cross section \cite{noda86}. 

The degeneracy of modes with different polarization and parity  is lifted in this kind of fibers due to the lack of azimuthal symmetry. Thus, modes $\mbox{LP}_{lm}^{ex}$, $\mbox{LP}_{lm}^{ey}$, $\mbox{LP}_{lm}^{ox}$, and $\mbox{LP}_{lm}^{oy}$ all have a different propagation constant \cite{wang05}.

In order to specify the effective refractive index of propagating modes in highly birefringent fibers, we use a simple model based on the treatment for circularly symmetric  step index fibers in the weakly guiding approximation \cite{snyder78}. Assuming that $n_0$ is the effective refractive index for a specific $\mbox{LP}_{lm}$ mode (with $l \ge1$)  in the circularly symmetric fiber, we model the refractive indices of the four unfolded modes as follows

\begin{eqnarray}
\label{eq:index}
n_{ey}=n_0,\qquad \qquad \qquad n_{oy}=n_0+\Delta_p, \nonumber \\ n_{ex}=n_0+\Delta, \qquad \qquad n_{ox}=n_0+\Delta+\Delta_p,
\end{eqnarray}

\noindent where $\Delta$ and $\Delta_p$ are the polarization and parity birefringences, respectively.  Note that for $l=0$ modes, odd parity is undefined, so that these relations simplify to

\begin{align}
\label{eq:index}
n_{y}&=n_0 \nonumber \\ n_{x}&=n_0+\Delta.
\end{align}

From these definitions the propagation constant for each mode can then be calculated, which allows us to explore the phasematching properties for all potential SFWM processes.

\section{Experimental implementation}

\begin{figure}
\includegraphics[width=7cm]{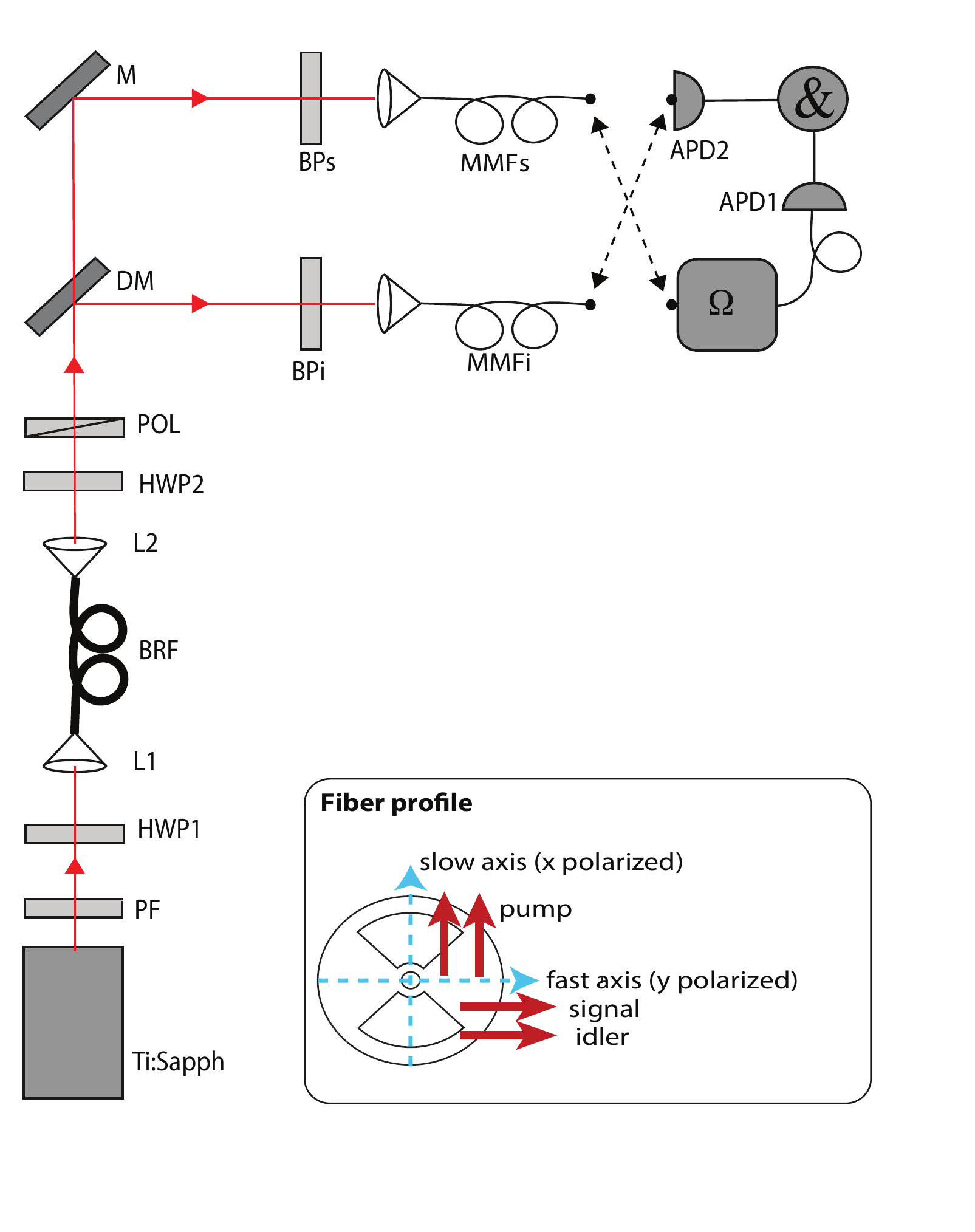}
\caption{\label{fig:setup} Experimental setup.}
\end{figure}

The SFWM source used in our experiments, sketched in  figure \ref{fig:setup},  is similar to the one used in our  papers Ref.~\cite{cruz14,cruz15} .  We employ as pump for the SFWM process a picosecond mode-locked Ti:sapphire laser  (76MHz repetition rate and $\sim0.5$nm bandwidth with a central wavelength which in our experiment is tuned from 690nm to 720nm).  The pump beam, filtered with a prism-based spectral bandpass filter (PF) with $\sim50$mW power, is coupled into a $14.5$cm length of bow-tie birefringent fiber using an aspheric lens with $8$mm focal length (L1); the polarization in the fiber is set parallel to the fiber's slow axis using a half-waveplate (HWP1).      We generate photon pairs in this fiber through cross-polarized SFWM so that the signal and idler photons are polarized parallel to the fiber's fast axis (see inset of Fig.~\ref{fig:setup}).  The photon pairs are out-coupled from the fiber using a second aspheric lens with $8$mm focal length (L2) and their polarization is set to horizontal using a second half wave plate (HWP2); a Glan-Thompson polarizer (POL)  reduces the remaining pump power by  a factor equal to the extinction ratio of $\sim 10^5$.     The photon pairs are frequency non-degenerate, emitted in spectral bands placed symmetrically around the pump;  they are split  using  a dichroic mirror (DM) followed by bandpass filters (BPs and BPi)  centered at the signal and idler photons for further pump suppression.   

Our primary aim is to measure the signal and idler emission spectra, as a function of the pump wavelength. For this purpose, we rely on spectrally-resolved, coincidence photon counting.  For a given pump central frequency, which we scan from $690$nm to $720$nm, we couple 
the idler  ($\lambda<\lambda_p$) and  signal  ($\lambda>\lambda_p$) photons into  separate multimode fibers, MMFi and MMFs.  For spectrally-resolved detection,  we employ a scanning grating-based monochromator ($\Omega$) which has been fitted with multimode fiber input and output ports.  While MMFs serves as input for $\Omega$,  with the output of $\Omega$ directed to a silicon avalanche photodiode (APD1),  fiber MMFi leads directly to a second avalanche photodiode (APD2).  We then monitor the coincidence rate at APD1 and APD2, with accidental counts subtracted, as a function of the central transmission frequency of $\Omega$.  In this manner we are able to measure the signal-photon spectrum in coincidence with the corresponding (spectrally-unresolved) idler photon.  Subsequently, we reverse the role of the two photons so as to measure the idler-photon spectrum in coincidence with the (spectrally-unresolved) signal photon. 

\begin{figure*}
\centering
\includegraphics[width=15cm]{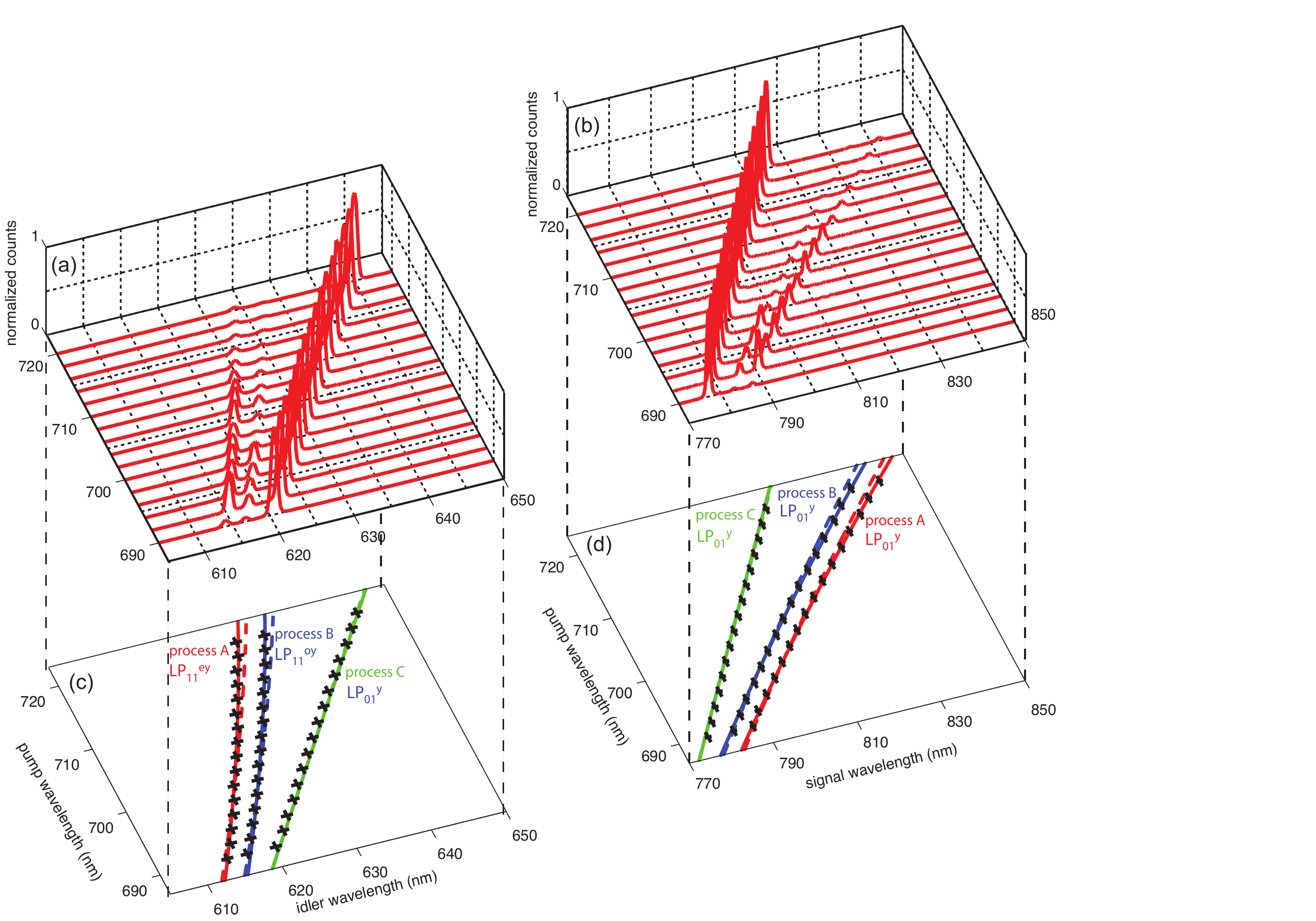}
\caption{\label{fig:PM_data} (a) and (b):   Measured SFWM spectra composed of three pairs of signal/idler energy conserving peaks, for a number of pump wavelengths within the range $690$ to $720$nm,  (c) and (d): SFWM wavelengths as a function of the pump wavelength for each of the three inferred processes; horizontal bars represent the signal/idler emission bandwidth while vertical bars indicate the pump bandwidth.  }
\end{figure*}

In Fig.~\ref{fig:PM_data} we show our experimental results.    In panel (a) we show a collection of idler-photon ($\lambda<\lambda_p$) spectra for a number of different pump wavelengths within the range $690$nm to $720$nm.   In panel (b) we show the corresponding signal-photon ($\lambda>\lambda_p$) spectra.     It is apparent from these plots, especially for the lower pump wavelengths, that the SFWM spectra are in the form of three pairs of energy-conserving peaks.   One of the pairs of peaks (the one which involves signal and idler wavelengths closest to the pump wavelength) leads to a  higher count rate (by about one order of magnitude, as compared to the other two pairs of peaks.    The three pairs of peaks are present for all of the pump wavelengths considered, although for the larger pump wavelengths two of them are much reduced in count rate; this is in all likelihood because the pump approaches the cutoff wavelengths for the LP$_{11}^{ex}$ and LP$_{11}^{ox}$ modes which serve as pumps for processes $A$ and $B$~\cite{comment}.  In panel (c) we show the location of each of the three idler-photon peaks, directly obtained from the peak maxima in Fig.~\ref{fig:PM_data}(a); for each of the experimental points, the horizontal bar indicates the width of the corresponding SFWM peak, while the vertical bar represents the pump bandwidth.    Panel (d) shows the corresponding signal-photon emission wavelengths as a function of the pump wavelength, obtained from panel (b).

It is to be expected that each of the pairs of peaks discussed in the previous paragraph is associated with a separate SFWM process, defined by a particular combination of transverse modes for the four waves involved.    In order to explore this, it is helpful to measure the transverse mode into which the signal and idler photons are emitted for each pair of peaks.   As reported in Ref.~\cite{cruz15}, we have employed a combination of spectrally- and spatially-resolved photon counting in order to measure the transverse spatial distribution associated with each of the six peaks in Fig.~\ref{fig:PM_data}; we have indicated in panels (c) and (d) of Fig.~\ref{fig:PM_data} the measured transverse mode next to each of the six curves.  Note that these transverse modes remain unchanged for the whole range of pump wavelengths considered.

\section{Analysis of experimental results}

Our measurements of frequency-resolved coincidence rates suggest  (for all pump frequencies considered within the range $690$nm to $720$nm) the presence of three separate SFWM processes, where each process contributes one pair of energy-conserving peaks.  Qualitatively, all pump wavelengths lead to the same behavior, except that the signal and idler wavelengths shift towards the IR as the pump wavelength is increased (see Fig.~\ref{fig:PM_data}).   An analysis of the phasematching properties in the fiber is needed in order to correctly identify the  SFWM process which gives rise to each measured pair of spectral peaks. Such an analysis demands knowledge of the chromatic dispersion and of the modes supported by the fiber.  With the information provided by the manufacturer and based on measurements of the spatial profile of the generated photons reported in Ref.~\cite{cruz15}, we were able to identify that our ``bow tie''  fiber supports two propagation modes  within the spectral range relevant in the experiment: $\mbox{LP}_{01}$ and $\mbox{LP}_{11}$. Nevertheless, as was explained in subsection \ref{modeTh}, these two modes in fact  unfold, in a birefringent fiber,  into six non-degenerate modes, which leads to $6^4=1296$ possible SFWM processes.   Restricting our attention to cross-polarized processes of the type $xx-yy$, as is done in our experiment, leaves $3^2 \times 3^2=81$ processes.     Additionally,  by invoking the conservation of OAM  and parity (see subsection \ref{cons}) it can be shown that only $15$ processes are actually viable.  Thus, each of the three pairs of spectral peaks observed in the experiment could result from one of these $15$ processes, subject to the phasematching constraint (see equation (\ref{eq:DK})).

Note that while the transverse modes in which each of the SFWM photons is generated, within each of the three peaks,  is known from our experimental measurements (as reported in Ref.~\cite{cruz15}), the transverse modes in which the pump waves propagate cannot easily be determined from the experiment.  Table~\ref{Tab}  lists all $15$ processes identified in the previous paragraph, classified into four mutually exclusive groups, where we have taken into account the measured transverse modes for the SFWM photons : i)  a single process which is compatible with pair of peaks $A$, ii) a single process which is compatible with pair of peaks $B$, iii) three processes which are compatible with pair of peaks $C$, and iv) those processes that are not compatible with any of the three pairs of peaks.

Note that the description used in this paper, including the unfolding of LP modes according to parity and polarization, is an improvement over the model used in our earlier paper~\cite{cruz14}. Note also that a reliable theoretical description of the fiber used in~\cite{cruz14} is complicated on account of its very thin ``inner cladding'' (the space between the stress-applying rods and the core):  different transverse modes extend to varying degrees outwards from the core, and some may reach the stress-applying rods.  This implies that a simple dispersion model based on the step index fiber with corrections due to birefringence parameters is no longer sufficient.
In this work we have used a fiber with the stress rods sufficiently removed from the core so that the  evanescent tails of the transverse modes are fully contained by the inner cladding.

\begin{table}
\begin{center}
\begin{tabular}{|c|c|c|c|c|}
\hline  & p1 & p2  & s($\lambda>\lambda_p$)& i($\lambda<\lambda_p$)   \\    
\hline       
\rowcolor[rgb]{0,1,1}{peaks A} & \textbf{01x}  & \textbf{11ex} & \textbf{01y} & \textbf{11ey}\\
\hline
\rowcolor[rgb]{1,0,1}{peaks B} & \textbf{01x} & \textbf{11ox} & \textbf{01y} & \textbf{11oy} \\
\hline
\cline{2-5}\cellcolor[rgb]{1,1,0}&\cellcolor[rgb]{1,1,0}\textbf{01x} & \cellcolor[rgb]{1,1,0} \textbf{01x}  & \cellcolor[rgb]{1,1,0} \textbf{01y} & \cellcolor[rgb]{1,1,0} \textbf{01y}\\
\cline{2-5}\cellcolor[rgb]{1,1,0}& \cellcolor[rgb]{1,1,0}11ex & \cellcolor[rgb]{1,1,0}11ex  & \cellcolor[rgb]{1,1,0}01y & \cellcolor[rgb]{1,1,0}01y\\
\cline{2-5}\multirow{-3}{*}{\cellcolor[rgb]{1,1,0}{peaks C}} &\cellcolor[rgb]{1,1,0}11ox& \cellcolor[rgb]{1,1,0}11ox & \cellcolor[rgb]{1,1,0}01y &\cellcolor[rgb]{1,1,0}01y\\
\hline
\multirow{10}{*} {others} & 01x & 01x & 11ey & 11ey\\
\cline{2-5}& 01x & 01x  & 11oy & 11oy\\
\cline{2-5}& 01x & 11ex  & 11ey & 01y\\
\cline{2-5}& 01x & 11ox  & 11oy & 01y\\
\cline{2-5}& 11ex & 11ex & 11ey & 11ey\\
\cline{2-5}& 11ex & 11ex  & 11oy & 11oy\\
\cline{2-5}& 11ex & 11ox  & 11ey & 11oy\\
\cline{2-5}& 11ex & 11ox & 11oy & 11ey\\
\cline{2-5}& 11ox & 11ox  & 11ey & 11ey\\
\cline{2-5}& 11ox & 11ox  & 11oy& 11oy\\
\hline
\end{tabular}
\caption{List of the 15 cross-polarized $xx-yy$  SFWM processes which conserve both OAM and parity.}\label{Tab}
\end{center}
\end{table}

\begin{figure}
\centering
\includegraphics[width=8.6cm]{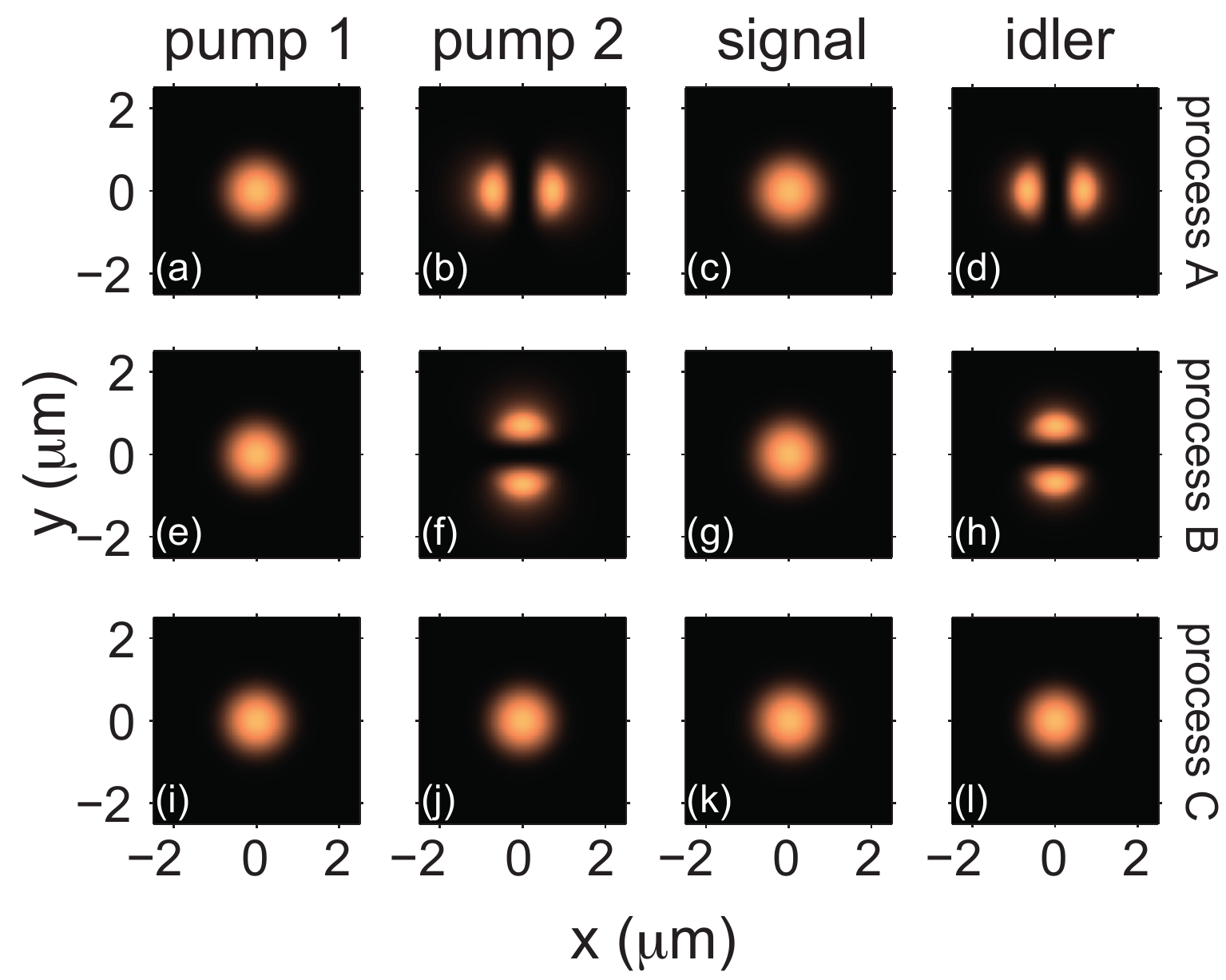}
\caption{\label{fig:allmodes} Transverse modes participating in each of processes $A$ (first row), $B$ (second row) and $C$ (third row).  The columns correspond to the four waves involved in the SFWM process: pump $1$, pump $2$, signal, and idler.}
\end{figure}

We have implemented a numerical strategy which accomplishes two tasks: it identifies i) the processes (from the groups in Table~\ref{Tab}), which optimize the simultaneous phasematching for all three pairs of peaks, and ii) the fiber parameters $\{r_0, NA, \Delta, \Delta_p\}$.  For the particular case in question the first task is trivial for two of the three pairs of peaks since there is a single process available. The search in the parameter space $\{r_0, NA, \Delta, \Delta_p\}$ is executed exploiting a genetic  algorithm (GA). In our GA, each individual is formed by the values of the four fiber parameters, and a fitness function (FF) is defined as $\Delta k_T=|\Delta k_A+\Delta k_B+\Delta k_C|$ (with  $\Delta k_A$/$\Delta k_B$/$\Delta k_C$ the
phase mismatch for process $A$/$B$/$C$, evaluated at the frequencies associated with each of the three pairs of peaks), with greater fitness associated with lower values of $\Delta k_T$.  A population of individuals evolve through successive generations of children created from the forebears by three mechanisms: i) the fittest ones (elite) are reproduced intact, ii) combination of parents, or crossover, and iii) the introduction of random mutations.  Note that in each generation step of the GA, all possible SFWM processes for each pair of peaks are tested retaining the combinations of processes which minimize the FF.

It is worth noting the FF can be defined for the three pairs of peaks obtained for a particular choice of pump wavelength or for collective data derived from $m_p$ pump wavelengths; in the latter case, the definition of  $\Delta k_T$ is adjusted so that $3 \times m_p$ $\Delta k's$ are added together, three for  each pump wavelength.   Also, for each pump wavelength, the algorithm may be run either: i) without specifying the signal and idler transverse modes (in which case $15$ mode combinations must be tested for each pair of peaks), or ii)  restricting to the experimentally-measured signal and idler trasnsverse modes.    Of course both alternatives should lead to the same result, which constitutes a useful self-consistency test (which is indeed fulfilled in our implementation).   Our algorithm leads to a number of solutions for the fiber parameters $\{r_0, NA, \Delta, \Delta_p\}$, all of them involving the processes shown in bold typeface in table~\ref{Tab}.  In Fig.~\ref{fig:allmodes} we  have plotted the  transverse modes that participate in each of the three proceses.

\begin{figure*}
\includegraphics[width=16cm]{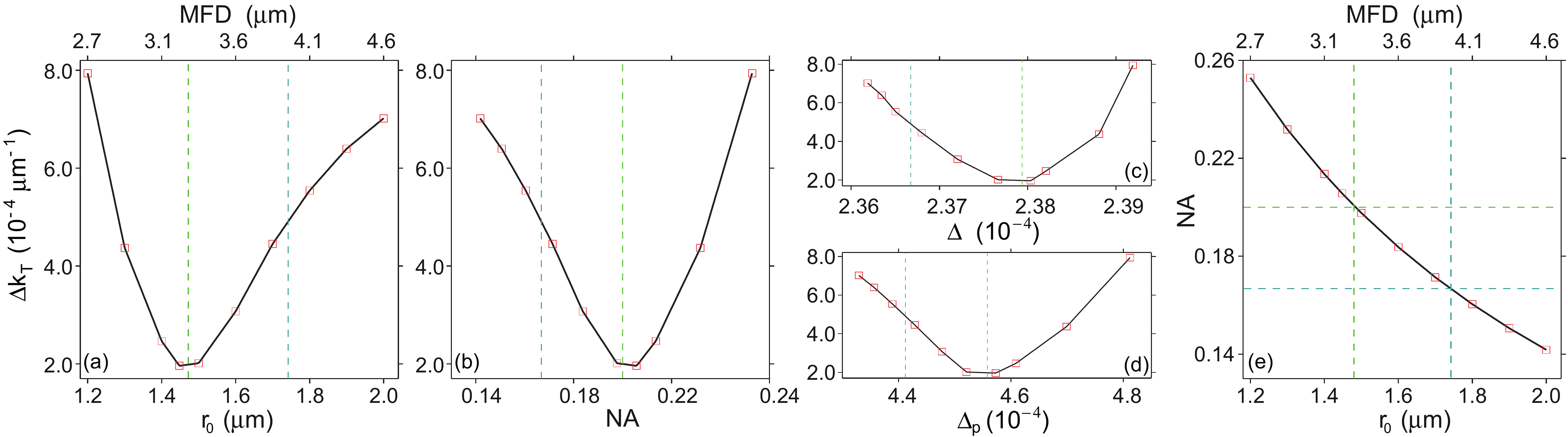}
\caption{\label{fig:NA_Radius} For the family of solutions that we obtain, we plot the fitness function $\Delta k_T$ vs core radius, in (a), vs numerical aperture, in (b), vs birefringence, in (c), and vs parity birefringence, in (d). Red dots: solutions obtained  from the GA;  black solid line:  interpolation.   In (e) we represent the family of solutions in the space formed by the core radius and the numerical aperture. Blue dotted line: data from manufacturer, green dotted line: leads to minimum fitness function.}
\end{figure*}

In addition to the determination of the specific processes present in our source, our GA yields a family of parameter sets  $\{r_{0}, NA, \Delta, \Delta_{p}\}$, each one leading to agreement to within $1.5$nm between measured and predicted peaks.  This parameter set family is represented with red squares in Figs.~\ref{fig:NA_Radius}(a)-(d), with the resulting values of $\Delta k_T$ plotted in the vertical axis, and where black lines represent interpolations which suggest a continuum of solutions; indeed, we verified that any combination of parameters from these curves yields good agreement with our experimental results.  In panels (a) and (e) we have included for convenience the core radius in the bottom and the mean field diameter (MFD) of the fundamental mode in the top.  Note that while the variation along this continuum of the birefringence parameters $\Delta$ and $\Delta_p$ is only slight, there is a much more significant variation in the radius and the numerical aperture.  Our family of solutions is represented in $\{r_{0}, NA\}$ space in Fig.~\ref{fig:NA_Radius}(e), indicating what may be regarded as a continuum of ``equivalent fibers'' all of which would yield a similar SFWM behavior.   Indeed, since the waveguide contribution to the overall fiber dispersion is proportional to the index contrast, i.e. to the $NA$, and inversely proportional to the core radius, the inverse relationship implied by Fig.~\ref{fig:NA_Radius}(e) agrees with intuition.   

Among the family of solutions shown in Fig.~\ref{fig:NA_Radius}, there are two which are of particular interest.   The first one, which involves the parameters 
$r_0=1.45\mu$m (implying  a mean filed diameter of  $MFD=3.27\mu$m),  $NA=0.20$, $\Delta=2.38\times 10^{-4}$, and $\Delta_p=4.57\times 10^{-4}$ is the one which minimizes the fitness function $\Delta k_T$.    The second one, which involves the parameters 
$r_0=1.742\mu$m (implying a mean field diameter of $MFD=4\mu$m), $NA=0.167$, $\Delta=2.37\times 10^{-4}$, and $\Delta_p=4.41\times 10^{-4}$, is compatible with the parameter values provided by the manufacturer (except for the parity birefringence $\Delta_p$, which is unspecified).   The first (second) of these identified solutions involves a maximum deviation between the experimental and theory peaks of $0.59$nm ($1.47$nm).     In figures~\ref{fig:PM_data}(c) and \ref{fig:PM_data}(d) we have shown phasematching curves (i.e. defined as the locus of pump and SFWM wavelengths for which $\Delta k=0$ for each of the three processes) overlapped with experimental points. Note the solid lines were plotted assuming the first solution from the previous paragraph, while dashed lines were plotted assuming the second solution. It is evident from this plot that: i) the two solutions identified in the previous paragraph yield essentially the same phasematching characteristics, and ii) there is excellent agreement between theory and experiment.   

A relevant aspect of our analysis of the experiment is that it could be exploited as a fiber characterization technique, applicable to different fiber geometries. As has been shown above, from a set of SFWM wavelength vs pump wavelength  experimental data, our genetic algorithm is capable of identifying a family of parameters $\{r,NA,\Delta,\Delta_p\}$ as candidates to describe the fiber used.   For the specific experimental situation studied here, on the one hand the two birefringence parameters $\Delta$ and $\Delta_p$ are relatively constant within the resulting family of solutions, and can be determined with an accuracy of $<1\%$ for $\Delta$ and  $<10\%$ for $\Delta_p$.  On the other hand, the radius and numerical aperture exhibit a particular inverse relationship between them, so that the independent determination of either of these two parameters would enable us to determine the remaining parameter.   

\begin{figure}
\centering
\includegraphics[width=8.6cm]{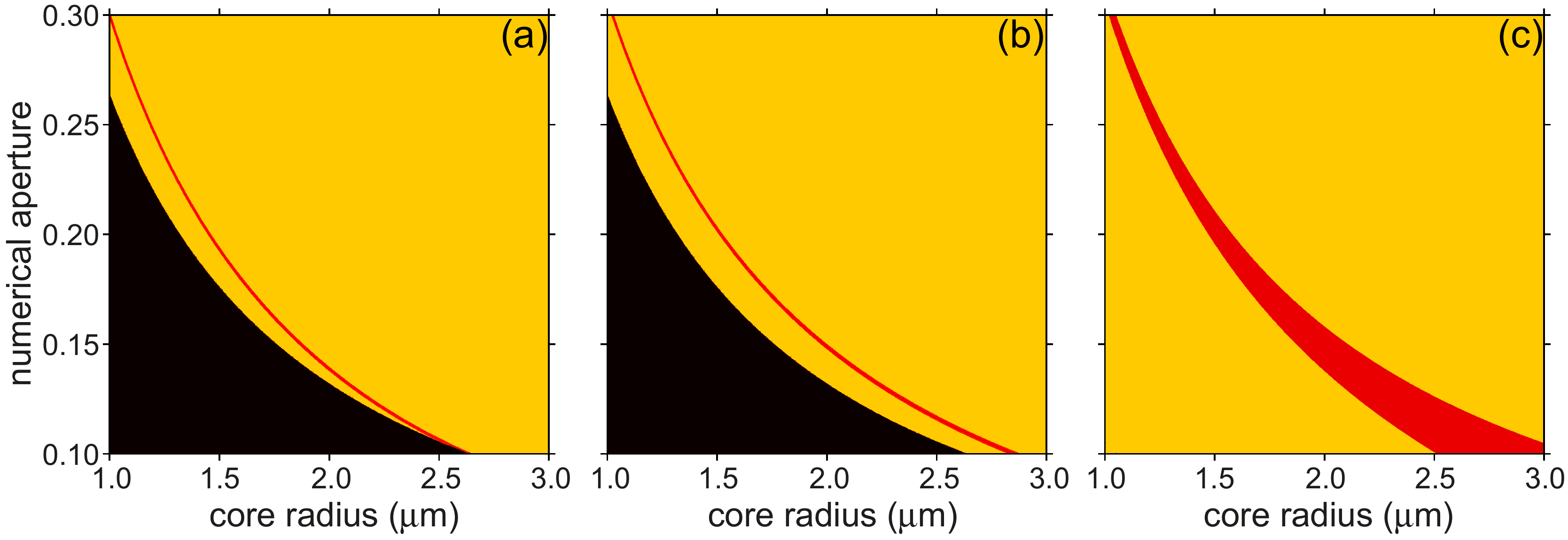}
\caption{\label{fig:DK_Proc} Phasemismatch function  $L |\Delta k|$ for (a) process A, (b) process B, and (c) process C, plotted as a function of the core radius and numerical aperture for fixed birefringence parameters.  Red regions indicate $L|\Delta k| \le 2 \pi$, yellow regions indicate $L |\Delta k| \ge 2 \pi$, while black regions indicate that the non-fundamental modes LP$_{11}$ are not supported. }
\end{figure}

In the context of the possible application of our experimental analysis  as a fiber characterization technique, it is interesting to compare the phasematching constraints for process C (for which all waves propagate in the fundamental mode), on the one hand, and for processes A and B (which involve non-fundamental modes), on the other hand. In figure~\ref{fig:DK_Proc} we show, in the space formed by the core radius and the numerical aperture, the phasematched region for which $| L\Delta k| \le 2 \pi$ for each of the three processes (red regions).   Clearly, processes A and B lead to considerably more acute phasematching constraints as compared to process C.    This implies that the existence of processes which involve higher-order modes improve the prospects of this analysis as a fiber characterization technique.   Neverthless, this fiber characterization technique could be applied in benefit of the considerable body of work which exists based on SFWM in birefringent fibers without the use of multiple transverse modes~\cite{smith09,zhou13,fang14,fan07}.

\section{Conclusions}

We have presented a theoretical and experimental study of a photon pair source based on the process of spontaneous four wave mixing (SFWM), in a situation where more than one transverse mode is supported, both for the pump and for the signal/idler photons.   We have based our work on a birefringent fiber, specifically on a ``bow-tie'' fiber, in which the propagation modes are well approximated by the linearly polarized (LP) family of modes.   We discuss that in the presence of birefringence, while the fundamental mode LP$_{01}$ unfolds into two non-degenerate modes associated with $x$ and $y$ polarizations, the LP$_{11}$ (along with all those with $l \ge 1$) unfolds into four non-degenerate modes labelled by the different combinations of $x$/$y$ polarizations and even/odd parity.   We have shown that 
within the linearly polarized approximation, among all possible SFWM processes defined by different combinations of transverse modes for the four participating waves, the departure from circular symmetry means that  both orbital angular momentum and parity must be conserved in order for a process to be  viable.  We discuss that the presence of multiple SFWM processes leads, in general, to hybrid two-photon entanglement  in frequency and transverse mode.

We present the results of an experiment in which we have measured, for a number of different pump wavelengths, the SFWM spectra in coincidence, which are composed of multiple (in this case three) pairs of energy-conserving peaks.    We use these experimental results, together with information about the transverse modes in which the signal and idler photons are generated, as input to a genetic algorithm which on the one hand enables us to match a particular SFWM process with each pair of energy-conserving peaks, and on the other hand permits us to characterize the fiber, i.e. to obtain numerical values for the fiber parameters.    We believe that these results will pave the way for further progress in the generation of photon pairs in optical fibers with hybrid entanglement in frequency and transverse mode.

\begin{acknowledgments}
This work was supported by CONACYT, M\'exico and by PAPIIT(UNAM) grant IN1050915.
\end{acknowledgments}


\begin{thebibliography}{99}

\bibitem {fiorentino02}M. Fiorentino, P. L. Voss, J. E. Sharping, and P. Kumar,  IEEE Photon. Technol. Lett. {\bf 14}, 983--985 (2002).

\bibitem{Burnham70} D. C. Burnham and D. L. Weinberg, Phys. Rev. Lett. {\bf 25}, 84--87 (1970).

\bibitem {randel11} S. Randel, R. Ryf, A. Sierra, P.J. Winzer, A. H. Gnauck, C. A. Bolle, R-J Essiambre, D. W. Peckham, A. McCurdy, and R. Lingle,  Opt. Express {\bf19}, 16697-16707 (2011).

\bibitem {chen13} Y. Chen, W. J. Wadsworth, and T. A. Birks,  Opt. Lett. {\bf38} 3747--3750 (2014).

\bibitem {shavrin13} I. Shavrin, S. Novotny, and H. Ludvigsen,  Opt. Express {\bf21}, 32141--32150 (2013).

\bibitem {mosley09} P. J. Mosley, A. Christ, A. Eckstein, and C. Silberhorn,  {\it Phys. Rev. Lett.} {\bf103} 233901 (2009).

\bibitem {christ09} A. Christ, K. Laiho, A. Eckstein, T. Lauckner, P. J. Mosley, and C. Silberhorn,  Phys. Rev. A {\bf80}, 033829 (2009).

\bibitem{kruse13} R. Kruse, F. Katzschmann, A. Christ, A. Schreiber, S. Wilhelm, K. Laiho, A. G\'abris, C. S. Hamilton, I. Jex, and C. Silberhorn, New J. Phys. {\bf15}, 083046 (2013).

\bibitem{karpinski09} M. Karpinski, C. Radzewicz, and K. Banaszek,  Appl. Phys. Lett. {\bf 94}, 181105 (2009).



%mosley09,christ09,kruse13,karpinski15

\bibitem{saleh09} M. F. Saleh, B. E. A. Saleh, and M. C. Teich,  Phys. Rev. A {\bf79} 053842 (2009).


\bibitem{neves09}see for example: L. Neves, G. Lima, A. Delgado, and C. Saavedra, Phys. Rev. A {\bf 80} 042322 (2009)
%\bibitem{kwiat97} P. G. Kwiat, J. Mod. Opt. {\bf 44}, 2173--2184 (1997).

\bibitem{cruz14}D. Cruz-Delgado, J. Monroy-Ruz, A. M. Barragan, E. Ortiz-Ricardo, H. Cruz-Ramirez, R. Ramirez-Alarcon, K. Garay-Palmett, and A. B. U'Ren,  Opt. Lett. {\bf39}, 3583-3586 (2014).


\bibitem {smith09} B. J. Smith, P. Mahou, O. Cohen, J. S. Lundeen, and I. A. Walmsley,  Opt. Express {\bf17}, 23589--23602 (2009).

\bibitem {zhou13} Q. Zhou, W. Zhang, T. Niu, S. Dong, Y. Huang, and J. Peng,  Eur. Phys. J. D {\bf67}, 202 (2013).

\bibitem {fang14} B. Fang, O. Cohen, and V. O. Lorenz, J. Opt. Soc. Am. B {\bf31} 277--281 (2014).

\bibitem {fan07} J. Fan and A. Migdall,  Opt. Express {\bf 15}, 2915--2920 (2007).

% SFWM theory

\bibitem{garay07} K. Garay-Palmett, H. McGuiness, O. Cohen, J. Lundeen, R. Rangel-Rojo, A.B. U'Ren, M. Raymer, C. McKinstrie, S. Radic and I.A. Walmsley, Opt. Exp, \textbf{15} 14870 (2007).

% SFWM engineering

\bibitem {cruz15} D. Cruz-Delgado, R. Ramirez-Alarcon, E. Ortiz-Ricardo, J. Monroy-Ruz, F. Dominguez-Serna, H. Cruz-Ramirez, K. Garay-Palmett, and A. B. U'Ren, submitted

% modes in birefirngent fibers


\bibitem{snyder78}A. W. Snyder and W. R. Young,  J. Opt. Soc. Am. {\bf68}, 297--309 (1978).

\bibitem{kim87}B. Y. Kim, J. N. Blake, S. Y. Huang, and H. J. Shaw,  Opt. Lett {\bf12}, 729--731 (1987).

% Conservation of OAM for SPDC

\bibitem{mair01} A. Mair, A. Vaziri, G. Weihs and A. Zeilinger Nature {\bf412}, 313-316 (2001)

\bibitem{arnaut00} H. H. Arnaut and G. A. Barbosa Phys. Rev. Lett. {\bf85}, 286 (2000)

\bibitem{barbosa07} G.A. Barbosa, Phys. Rev. A {\bf76} 033821 (2007)

\bibitem{feng08} S. Feng and P. Kumar,  Phys. Rev. Lett. {\bf101}, 163602 (2008)

\bibitem{barbosa09} G.A. Barbosa Phys. Rev. Lett. {\bf103}, 149303 (2009)



% Conservation of parity for SPDC

\bibitem{walborn05} S. P. Walborn, S. Padua, and C. H. Monken
Phys. Rev. A {\bf71}, 053812 (2005)

% Conservation of OAM - parity in SPDC waveguides 

\bibitem{bharadwaj15} D. Bharadwaj, K. Thyagarajan, M. Karpinski, and K. Banaszek
Phys. Rev. A {\bf91}, 033824 (2015)

\bibitem {mosley09} P. J. Mosley, A. Christ, A. Eckstein, and C. Silberhorn, Phys. Rev. Lett. {\bf 103}, 233901 (2009).



% OAM - parity conservation in classical processes

\bibitem{lin81} See for example: C. Lin, and M. A. Bosch,  Appl. Phys. Lett.Ê {\bf38}, 479 (1981);  C. Lesvigne, V. Couderc, A. Tonello, P. Leproux, A. BarthŽlŽmy, S. Lacroix, F. Druon, P. Blandin, M. Hanna, and P. Georges,  Opt. Lett. {\bf32}, 2173-2175 (2007);  J. Cheng, M.E.V. Petersen, K Charan, K. Wang, C. Xu, L. Gruner-Nielsen, and D. Jakobsen, Appl. Phys. Lett. {\bf101}, 161106 (2012)



% overlap integrals classical FWM
\bibitem{poletti08} F. Poletti and p. Horak,  J. Opt. Soc. Am. B  {\bf25}, 1645 (2008)

\bibitem{ding14} Y. Ding, J. Xu, H. Ou, and C. Peucheret, Opt. Express {\bf22}, 127 (2014)





\bibitem {meyer13} E. Meyer-Scott, V. Roy, J-P. Bourgoin, B. L. Higgins, L. K. Shalm, and T. Jennewein,  Opt. Express {\bf21}, 6205--6212 (2013).

\bibitem {lin07} Q. Lin, F. Yaman, and G. P. Agrawal,  Phys. Rev. A {\bf75}, 023803 (2007).




\bibitem {noda86} J. Noda, K. Okamoto, and Y. Sasaki,  J. Lightwave Technol. {\bf4}, 1071--1089 (1986).

\bibitem {wang05} Z. Wang, J. Ju, and W. Jin,  Opt. Express {\bf13}, 4350-7 (2005).



\bibitem{comment} Note that while the cutoff wavelength for the LP$_{11}$ mode is $759$nm, our approximate dispersion model cannot predict the cutoff wavelengths for the LP$_{11}^{ex}$ and LP$_{11}^{ox}$ modes.



\end{thebibliography}
\end{document}